# Social Interactions Mediated by the Internet and the Big-Five: a Cross-Country Analysis

Andrea Mercado[1], Alethia Hume[1], Ivanno Bison[2], Fausto Giunchiglia[2], Amarsanaa Ganbold[3] and Luca Cernuzzi[1]

[1] *DEI - Universidad Católica Nuestra Señora de la Asunción - Paraguay*
[2] *University of Trento - Italy*
[3] *National University of Mongolia*

## Abstract
This study analyzes the possible relationship between personality traits, in terms of Big Five (extraversion, agreeableness, responsibility, emotional stability and openness to experience), and social interactions mediated by digital platforms in different socioeconomic and cultural contexts. We considered data from a questionnaire and the experience of using a chatbot, as a mean of requesting and offering help, with students from 4 universities: University of Trento (Italy), the National University of Mongolia, the School of Economics of London (United Kingdom) and the Universidad Católica Nuestra Señora de la Asunción (Paraguay). The main findings confirm that personality traits may influence social interactions and active participation in groups. Therefore, they should be taken into account to enrich the recommendation of matching algorithms between people who ask for help and people who could respond not only on the basis of their knowledge and skills.

## 1. Introduction

Among other diversity dimensions, personality traits may play a relevant role in social interactions mediated by technological platforms [1, 2, 3, 4, 5, 6, 7, 8]. Moreover, the socio-cultural context may influence the diversity [9]. For this study, we analyze 4 pilot experiments that were carried out in parallel with university students in different socioeconomic and cultural contexts; that is, University of Trento (Italy), the National University of Mongolia, the School of Economics of London (United Kingdom) and the Universidad Católica Nuestra Señora de la Asunción (Paraguay), using a self-reported questionnaire to begin modeling and analyzing diversity among students based on their social practices, competencies, knowledge and motivations. One of the survey dimensions focused on personality traits. Despite some criticism [13], we adopted the widely used Big-Five model [10]: Extraversion (E), Agreeableness (A), Conscientiousness (C), Neuroticism (N) and Openness (O). Complementarity, a Chatbot application allows participants social interactions requesting and offering help, represented as questions and answers in the application. Thus, the main objective of the research is to analyze the role played by personality in social interaction mediated by a Chatbot. This could inform machine algorithms based on artificial intelligence for recommending persons that could offer better help.

## 2. Data from the pilots

The full data collection process was identically applied in all the four pilot sites. In order to standardize the tools and experiences, translation to English (for a normalization of the values) and localization work was necessary to adapt them to the sociolinguistic skills of each site. The organizational details, as well as the ethical and legal aspects, are described in [11].

Participants have been recruited through email invitations and classified according to the area of study: STEM or No-STEM. Finally, the collected data are anonymized by each institution and made available to WeNet collaborators to inform machine learning algorithms able to enhance interactions between students and contribute to the "Diversity Model". Among the almost 13 thousand responses to the survey from which about 8500 complete psychosocial profiles, we invited the target population to participate in the Chatbot experience. Users generate both questions and answers to queries through interaction with other students of the same institution and even provide suggestions on a topic of interest or simply comments. The participation was voluntary, subject to the availability and interests of the users. The following table shows the participation in the Chatbot experience in the different sites.

**Table 1**
Number of participants (P), answers (A) and questions (Q), disaggregated by area of study and sex.

|         | UNITN | | | LSE | | | NUM | | | UC | | |
|---------|---|---|---|---|---|---|---|---|---|---|---|---|
|         | P | Q | A | P | Q | A | P | Q | A | P | Q | A |
| Male    | 14 | 78 | 443 | 5 | 15 | 117 | 8 | 56 | 432 | 10 | 85 | 234 |
| Female  | 28 | 265 | 593 | 38 | 233 | 608 | 29 | 497 | 2481 | 10 | 119 | 310 |
| STEM    | 16 | 100 | 460 | 7 | 120 | 442 | 21 | 496 | 2379 | 15 | 73 | 196 |
| No-STEM | 26 | 243 | 576 | 36 | 128 | 283 | 16 | 57 | 534 | 5 | 131 | 348 |
| **Total** | **42** | **343** | **1036** | **43** | **248** | **725** | **37** | **553** | **2913** | **20** | **204** | **544** |

## 3. Analysis of results

The data obtained during the experiment were analyzed in relation to the personality traits of the Chatbot users according to the Big-Five taking into account the length of questions and answers input by the participants, and the possible effect of other sociodemographic variables (sex, area of study, and site of the pilot). For the analysis, the Spearman's rank correlation test was used, as in previous work [1], but this time in combination with multinomial regression.

In Table 2, the correlation between the length of questions and answers and Big-Five is shown, the analysis is done both by pilots and in total for the whole dataset. By looking into this data for each institution some correlations can be found. However, it can also be noted that when analyzing the dataset in this more fragmented way, the values and signs of correlations sometimes change. This can be due to the size and composition of the samples, or other elements, like the translations, that can make the error more significant in the predictions based on these results. In this sense, it is also reasonable to assume that personality characteristics, and therefore the effects they may have on users behavior, do not change across cultures. Hence, the correlations are also analyzed over the total number of users in the entire dataset. Thus, in general a negative correlation can be identified regarding the length of questions with Neuroticism; while the length of answers shows positive correlations with Extraversion, Agreeableness and Openness, and negative ones with Conscientiousness and Neuroticism.

**Table 2**
**Spearman correlation between logarithmic length of questions and answers and Big-Five.**

|  | Total | | LSE | | NUM | | UC | | UNITN | |
|---|---|---|---|---|---|---|---|---|---|---|
|  | Corr. | p. | Corr. | p. | Corr. | p. | Corr. | p. | Corr. | p. |
| *Question* | | | | | | | | | | |
| E | 0.032 | 0.249 | **-0.109*** | **0.084** | **0.072*** | **0.088** | -0.013 | 0.852 | 0.044 | 0.456 |
| A | -0.041 | 0.139 | -0.033 | 0.599 | 0.043 | 0.310 | **-0.314*** | **0.000** | -0.033 | 0.574 |
| C | 0.020 | 0.474 | 0.078 | 0.217 | -0.009 | 0.835 | **-0.203*** | **0.003** | **0.104*** | **0.081** |
| N | **-0.155*** | **0.000** | -0.028 | 0.660 | **-0.111*** | **0.009** | **-0.290*** | **0.000** | -0.088 | 0.138 |
| O | -0.019 | 0.505 | -0.008 | 0.903 | -0.022 | 0.609 | 0.024 | 0.734 | -0.090 | 0.130 |
| Events | 1306 | | 255 | | 558 | | 207 | | 286 | |
| *Answer* | | | | | | | | | | |
| E | **0.0930*** | **0.000** | -0.053 | 0.150 | -0.017 | 0.330 | **0.0929*** | **0.021** | **0.1713*** | **0.000** |
| A | **0.1216*** | **0.000** | **0.1840*** | **0.000** | 0.004 | 0.803 | **0.0671*** | **0.097** | 0.036 | 0.231 |
| C | **-0.0375*** | **0.005** | 0.005 | 0.891 | **-0.0340*** | **0.053** | **0.0712*** | **0.078** | 0.048 | 0.114 |
| N | **-0.0486*** | **0.000** | 0.007 | 0.845 | **-0.0626*** | **0.000** | **-0.1365*** | **0.001** | **-0.0895*** | **0.003** |
| O | **0.0969*** | **0.000** | 0.033 | 0.370 | **0.0722*** | **0.000** | 0.002 | 0.968 | **0.0737*** | **0.015** |
| Events | 5688 | | 750 | | 3223 | | 614 | | 1101 | |

**Table 3**
**Multilevel multinomial linear regression of question-and-answer length.**

|  |  | Questions | | Answers | |
|---|---|---|---|---|---|
|  |  | Coef. | p. | Coef. | p. |
|  | Length of question |  |  | -6.7488 | 0.463 |
| Big-five | E | 0.0968 | 0.134 | 0.0752 | 0.78 |
|  | A | **-0.2232** | **0.049** | **-1.4582** | **0.001** |
|  | C | -0.0265 | 0.727 | 0.3643 | 0.26 |
|  | N | -0.0522 | 0.452 | 0.3468 | 0.242 |
|  | O | **0.1441** | **0.066** | **-0.5899** | **0.051** |
| Big-five * Length of question | E*lques |  |  | -0.0541 | 0.343 |
|  | A*lques |  |  | **0.3586** | **0** |
|  | C*lques |  |  | -0.0734 | 0.288 |
|  | N*lques |  |  | **-0.1432** | **0.026** |
|  | O*lques |  |  | **0.2111** | **0.001** |
|  | pilot (Ref. UNITN) |  |  |  |  |
|  | LSE | -6.5271 | 0.323 | **23.5338** | **0.003** |
|  | NUM | -1.4537 | 0.823 | **-21.096** | **0.008** |
|  | UC | 8.1407 | 0.278 | -4.06 | 0.649 |
|  | Sex (Ref. Male) |  |  |  |  |
|  | Female | -3.9917 | 0.487 | 0.1042 | 0.988 |
|  | Dep. (Ref. STEM) |  |  |  |  |
|  | No-STEM | 0.9058 | 0.867 | -5.2058 | 0.434 |
|  | Cons | 72.5146 | 0.000 | 84.7309 | 0.043 |
|  | Obs. | 115 |  | 105 |  |
|  | Events | 1318 |  | 5386 |  |

| | | |
|---|---|---|
| Wald chi2 | 16.54 | 397.98 |
| p. | 0.0853 | 0.0000 |

To further analyze dimensions that could influence the level of participation of participants, Table 3 shows a Multilevel multinomial linear regression of questions and answer length. As it can be seen sociodemographic variables, like sex and area of study (i.e., STEM, NO-STEM) appear to have no effect in predicting questions and answers lengths. These results also seem to confirm that, ceteris-paribus of personality traits and sociodemographic characters, a possible effect of cultural differences between the pilots only for the answers and not for the questions length. But these differences in answers seem to be due to English translation and not real cultural differences..

On the other hand, it is confirmed that personality does have a statistically significant effect. However, in order to better dimension this effect, the length of the answer is considered also with respect to the length of the question. In other words, when faced with banal, short questions such as "How are you?", we cannot expect very long answers, regardless of the personality of the respondents. Whereas, when faced with questions that give room for further elaboration of the answer, we can expect the effects of personality traits to emerge. In this sense, the results only show a positive effect of the personality traits Agreeableness and Openness, and a negative effect of Neuroticism, which affect the richness of the response.

Finally, Figure 1 shows projections of linear predicted answer lengths by Agreeableness and Openness and question length. That is, as the question becomes more articulate (more characters) so will the answer for people with high Agreeableness and Openness.

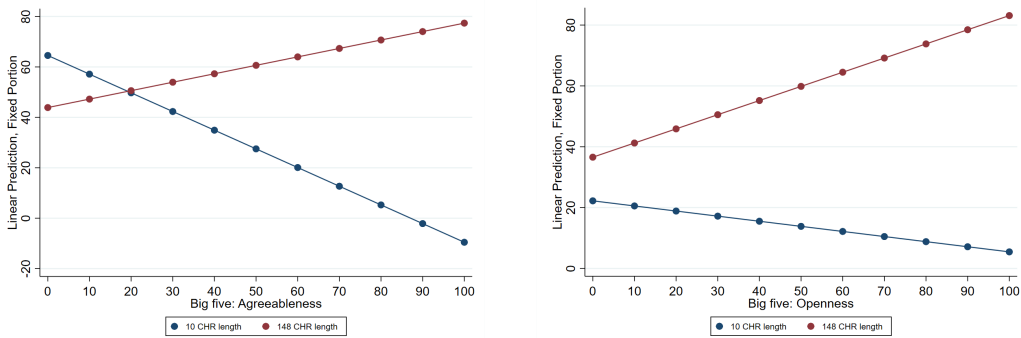

**Figure 1**: Predicted answer characters by Big-five (Agreeableness and Openness) and question length.

## 4. Discussion and conclusions

We have found that some personality traits of participants, modeled according to the Big-Five (such as Agreeableness and Openness to experience), influence the way they request help and/or contribute to other users through a Chatbot application. Moreover, other elements like sociodemographic variables appear to have no effect in predicting questions and answers lengths. With regard to potential cultural differences affecting response length, the sample is too small for a definitive conclusion. Further analysis may shed more light on the role of personality in characterizing diversity as a factor to improve Internet-mediated social interactions in different contexts.

## Acknowledgements

This research has received funding from the European Union's Horizon 2020 FET Proactive project "WeNet: Internet of us", Grant Agreement No: 823783.